\documentclass[twocolumn]{article}
\usepackage{graphicx}
\usepackage{float}
\usepackage{mathtools}
\usepackage{algorithm}
\usepackage{tabularx} 
\usepackage{array}
\usepackage{balance}
\usepackage{xspace}
\newcolumntype{C}[1]{>{\centering\let\newline\arraybackslash\hspace{0pt}}m{#1}}
\usepackage{graphicx} 
\usepackage{booktabs}
\usepackage{float}
\usepackage{amsmath,amssymb,amsthm,amsfonts,xcolor, color,stmaryrd}
\usepackage{enumerate}
\usepackage[utf8]{inputenc}
\usepackage[english]{babel}
\usepackage{tikz}
\usetikzlibrary{matrix, shapes, arrows, positioning, chains}
\usepackage{ragged2e}
\usepackage{etoolbox}
\usepackage{lipsum}
\usepackage[margin=0.9in]{geometry}
\usepackage{cuted}
\usepackage{hyperref}
\usepackage{siunitx}
\usepackage{authblk}
\usepackage[english]{babel}
\usepackage[numbers,sort&compress]{natbib}
\usepackage{balance}
\usepackage{amsmath}
\date{}
\usepackage{textgreek} 
\apptocmd{\frame}{}{\justifying}{} 
\usepackage{amsmath,amssymb,amsfonts}

\title{\textbf{Cosmic Expansion in $f(Q,C)$ Gravity: Observational Constraints and Geometric Diagnostics}}

\author[1]{Rajalaxmi Jena \thanks{rajalaxmi.jena2024@vitstudent.ac.in}} 
\author[1]{Sankarsan Tarai \thanks{sankarsan.tarai@vit.ac.in}} 

\affil[1]{\it Department of Mathematics, School of Advanced Sciences, Vellore Institute of Technology, Chennai-600127, India}

\begin{document}
\maketitle

\begin{abstract}
\noindent

We investigate the late-time expansion history of the Universe within the framework of $f(Q, C)$ gravity by adopting a linear parametrization of the deceleration parameter $q(z) = \alpha+\beta H$, where $Q$ denotes the non-metricity scalar and $C$ represents the associated boundary term satisfying the geometrical relation $\mathring{R}= Q+C$. The adopted parametrization provides a phenomenological description of the cosmic expansion while retaining analytical tractability, allowing the transition from decelerated to accelerated expansion to be examined directly. An exact integration of the corresponding differential equation yields the closed-form Hubble function, $H(z)=\frac{1}{\beta\left[F_0-\ln(1+z)\right]},$ thereby providing an explicit description of the expansion history in terms of redshift. Here $F=\frac{1}{\beta}\sqrt{2\beta t+b}$ and $F_0$ specifies the value of $F$ at the present epoch. The model parameters are constrained using recent Cosmic Chronometer (CC) measurements, DESI DR2 Baryon Acoustic Oscillation (BAO) observations, Pantheon+SH0ES Type Ia supernovae, and the Union3 observations through Bayesian Markov Chain Monte Carlo (MCMC) analysis. Individual and combined datasets are compared to assess the model's consistency, while its cosmic evolution is examined using the deceleration, the equation-of-state parameter, the state-finder pair ${r,s}$, and the $Om$ diagnostic. 
\end{abstract}

\medskip
\noindent\textbf{Keywords:} $f(Q,C)$ gravity; non-metricity; boundary term; deceleration parameter; dark energy.

\section{Introduction}

Following Einstein’s formulation of General Relativity (GR) in 1916 \cite{Einstein1922}, a decelerating Universe remained the prevailing cosmological expectation for much of the twentieth century. This view was grounded in the Friedmann equations, which describe the expansion of the Universe under the influence of its matter and radiation content \cite{Friedman1922}. In a Universe dominated by these components, gravitational attraction was expected to decelerate cosmic expansion. However, the discovery of late-time acceleration challenged this conventional picture, marking a significant turning point in modern cosmology \cite{Allen2004, Tegmark2004, Riess1998, Perlmutter1997, Perlmutter1998}. This departure from the conventional understanding prompted the exploration of alternative cosmological models and extensions of standard gravitational theory to explain the observed acceleration. The evidence in favor of the late-time accelerated expansion of the Universe came originally from Type Ia supernovae \cite{Clifton_2012, Riess2019} and has subsequently been confirmed by several other independent cosmological observations, most notably by the observation of the Cosmic Microwave Background and Baryon Acoustic Oscillations in the large-scale distribution of galaxies. In the standard cosmological description, this accelerated phase is accounted for by introducing dark energy (DE), an additional component of the cosmic energy budget whose sufficiently negative pressure gives rise to accelerated expansion \cite{Nojiri2007}. Among the proposed explanations for cosmic acceleration, dark energy (DE) remains one of the most extensively studied approaches to describing the late-time evolution of the Universe. Dark matter (DM) is introduced to explain the formation of large-scale cosmic structures and gravitational effects observed across galactic and cosmological scales. Together, the dark components constitute the dominant share of the Universe’s present cosmic energy budget, while ordinary baryonic matter accounts for only a small fraction.\par\bigskip

The cosmological constant $\Lambda$ provides the simplest and most successful explanation of dark energy, characterized by a constant vacuum-energy density and an equation-of-state parameter $\omega=-1$. Together with cold dark matter, it forms the basis of the standard $\Lambda$CDM cosmological model \cite{Komatsu2011}. Despite its remarkable agreement with observations of cosmic expansion and structure formation, the fundamental physical nature of both dark components remains unresolved. The cosmological constant also presents substantial theoretical difficulties, most notably the fine-tuning problem, arising from the vast disparity between the observed vacuum-energy density and theoretical estimates \cite{Copeland2006}. Despite having different evolutionary histories, dark matter and dark energy show similar energy densities in the relatively recent cosmic past, raising serious concerns about the coincidence problem. Potential differences between the estimated ages of the Universe's oldest observed astrophysical objects and the anticipated age of the Universe present a related problem. These constraints, along with the absence of a satisfying basic explanation for dark energy, encourage more research into cosmological models outside of the conventional $\Lambda$CDM framework.\par\bigskip

An alternative approach to explaining cosmic acceleration is to extend the gravitational framework itself, allowing late-time accelerated expansion to emerge from modified gravitational dynamics without introducing an additional exotic energy component. Such extensions generalize the Einstein-Hilbert description by adopting more general functions of curvature or incorporating other geometric invariants into the gravitational action. This approach has given rise to several modified gravity theories (MGTs), notably $f(R)$ gravity, which generalizes the gravitational Lagrangian through the Ricci scalar $R$ \cite{NOJIRI2011}; $f(T)$ gravity, formulated in terms of the torsion scalar $T$ \cite{Bengochea2009}; and $f(G)$ gravity, which incorporates the Gauss-Bonnet invariant $G$ \cite{Nojiri_2007, DEFELICE20091}. More general formulations include $f(R, T)$ gravity, which couples the Ricci scalar to the trace $T$ of the energy-momentum tensor \cite{Harko2011, Myrzakulov2012}, and $f(R, G)$ gravity, incorporating both Ricci curvature and Gauss–Bonnet contributions \cite{Bamba_2010}. Alongside these developments, symmetric teleparallel formulations have received increasing attention, particularly $f(Q)$ gravity, in which nonmetricity, represented by the scalar $Q$, provides the geometric description of gravitation in the absence of curvature and torsion \cite{Nester1999, Lazkoz2019, Jim2020, Anagnostopoulos_2021, Myrzakulov2022, Narawade_2024, Banerjee2021}. Its extension to $f(Q,T)$ gravity introduces an additional dependence on the trace of the energy-momentum tensor, allowing a coupling between nonmetricity and matter \cite{Xu_2020, SHARIF2024}. Collectively, these approaches broaden the study of gravitation beyond curvature-based formulations to encompass torsion and nonmetricity, offering alternative geometric foundations for investigating cosmological dynamics beyond GR.\par\bigskip

The theories of $f(Q, C)$, which extend symmetric teleparallel gravity, introduce an additional boundary term $C$ into the gravitational action, which allows the theories to give a more general geometric description of the late-time cosmic acceleration. A generic dependence on these scalars could lead to additional gravitational degrees of freedom, yielding richer cosmological dynamics. The scalar $Q$ originates from the nonmetricity tensor that describes the non-preservation of the metric under affine parallel transport, while $C$ is the total-divergence term that connects $Q$ to the Levi-Civita Ricci scalar through $\mathring{R}=Q+C$.\par\bigskip

The action governing the dynamics of $f(Q, C)$ gravity takes the form
\begin{equation}
S = \int\left[\frac{1}{2\kappa}f(Q,C)+\mathcal{L}_m\right]\sqrt{-g}\,d^4x,
\label{eq:action}
\end{equation}

The function $f(Q, C)$ determines the dependence of the gravitational Lagrangian on the non-metricity scalar $Q$ and the boundary term $C$. The constant $\kappa=8\pi G/c^{4}$ sets the gravitational coupling, whereas $\mathcal{L}_{m}$ specifies the matter contribution to the action. The invariant volume element is expressed through the tetrad determinant, $e=\det(e^{A}_{\mu})=\sqrt{-g}$, where $g$ is the determinant of the metric tensor.\par\bigskip

The gravitational field equations follow from requiring the action to be stationary under variations of the metric

\begin{strip}
\noindent
\rule{\dimexpr(\textwidth-\columnsep)/2\relax}{0.4pt}%
\rule{0.4pt}{5pt}\par
\vspace{6pt}
\begin{equation}
\kappa T_{\mu\nu} = -\frac{f}{2}g_{\mu\nu}+\frac{2}{\sqrt{-g}}\partial_\lambda \left(\sqrt{-g}f_Q P^{\lambda}_{\ \mu\nu}\right) + \left(P_{\mu\alpha\beta}Q_\nu^{\ \alpha\beta}-2P_{\alpha\beta\nu}Q^{\alpha\beta}_{\ \ \mu}\right)f_Q 
+\left[\frac{C}{2}g_{\mu\nu}-\mathring\nabla_\mu\mathring\nabla_\nu+g_{\mu\nu}\mathring\nabla^\alpha\mathring\nabla_\alpha-2P^{\lambda}_{\ \mu\nu}\partial_\lambda\right]f_C
\label{eq:fieldeq}
\end{equation}
\vspace{6pt}
\noindent\hfill
\rule[-5pt]{0.4pt}{5.4pt}%
\rule{\dimexpr(\textwidth-\columnsep)/2\relax}{0.4pt}\par
\end{strip}

The partial derivatives of the gravitational Lagrangian $f(Q,C)$ with respect to the non-metricity scalar $Q$ and the boundary term $C$ describe its dependence on these geometric quantities and are defined by

\begin{equation}
f_Q \equiv \frac{\partial f(Q,C)}{\partial Q},
\qquad
f_C \equiv \frac{\partial f(Q,C)}{\partial C}.
\end{equation}

The development of $f(Q, C)$ gravity has been aided by the necessity to connect several geometric models of gravitation and explain cosmic acceleration, explored in recent studies \cite{De_2024,Gadbail_2023}.  Samaddar et al. \cite{Amit2024} studied rip cosmology in $f(Q, C)$ gravity, analyzing the physical parameters driving Little Rip, Big Rip, and Pseudo Rip situations. In addition, \cite{Maurya2024} investigated quintessence-like cosmological evolution within $f(Q,B)$ gravity by adopting the quadratic form $f(Q,B)=\alpha Q^{2}+\beta B^{2}$.By contrasting the term $B$ with the Gibbons–Hawking–York boundary term of general relativity, Capozziello et al.\cite{Capozziello_2023} investigated the function of boundary contributions in $f(Q, B)$ symmetric teleparallel gravity.
\par\bigskip

Motivated by this framework \cite{Myrzakulov_2025, Myrzakulov2025}, we consider the functional form $f(Q, C)=\gamma_{1}Q^{n}+\gamma_{2}C$, where $\gamma_{1}$ and $\gamma_{2}$ are constant model parameters, and $n>1$ is chosen to explore nonlinear modifications in the nonmetricity sector. Since the term $\gamma_{2}C$ contributes only a boundary term to the action, the modifications to the bulk cosmological equations arise from $\gamma_{1}Q^{n}$. This ansatz therefore provides a setting for investigating the influence of nonlinear nonmetricity on the expansion history, including the transition from an early decelerating epoch to the present accelerated phase. Related studies have explored other nonlinear gravitational Lagrangians and their cosmological implications. In \cite{Rana2024}, the model $f(Q)=-Q+\alpha Q^{n}$ was examined using a dynamical-systems approach.Pradhan et al. \cite{Pradhan2024} investigated dark-energy cosmology in a spatially flat FLRW spacetime within \(f(Q, C)\) gravity. Adopting the form \(f(Q, C)=Q+\alpha C^{2}\), they examined the resulting cosmological evolution under observational constraints.\par\bigskip

Observations indicate that the Universe evolved from an earlier decelerating phase to late-time accelerated expansion. The deceleration parameter $q$ provides a direct measure of this evolution, which can be investigated using either parametric or non-parametric methods. Parametric approaches constrain an assumed form of $q$ and constrain its parameters against observational data, offering a concise, albeit model-dependent, description of cosmic expansion. \cite{Turner2002, Riess2004}. In contrast, non-parametric methods allow greater flexibility by avoiding a predefined functional form, although the robustness of their reconstruction depends on the quality and sampling of observational data.\cite{Shafieloo2007, Holsclaw2011}. A suitable parametrization of $q$ therefore provides a practical tool for studying the expansion history.\par\bigskip

In the present analysis, we employ a linear cosmological parametrization to characterize the expansion history within a simple, physically interpretable framework. The model parameters are constrained using observational data, and the resulting expansion history is subsequently assessed against independent cosmological datasets. The reconstructed evolution is then investigated using the Hubble parameter, the deceleration parameter, the effective equation-of-state parameter, and higher-order kinematic diagnostics. Particular emphasis is placed on the transition from decelerated to accelerated expansion and on the behaviour of the model across low and intermediate redshifts.\par\bigskip

The remainder of this paper is organized as follows. Section 2 outlines the theoretical foundations of $f(Q, C)$ gravity and derives the field equations for an FLRW spacetime. Section 3 introduces the adopted parametrization of the deceleration parameter and obtains the corresponding expression for the Hubble parameter. Section 4 presents the observational analysis, employing Bayesian inference to constrain the model parameters using cosmological datasets. Section 5 examines the cosmographic evolution of the model through its kinematic parameters and their implications for the expansion history. Finally, Section 6 discusses the principal findings and presents the conclusions of the theoretical and observational analyses.\par\bigskip

\section{Metric and field equation}

To examine the cosmological evolution of the Universe, we adopt a homogeneous and isotropic background spacetime. Under the additional assumption of spatial flatness on cosmological scales, this geometry is described by the Friedmann-Lemaître-Robertson-Walker (FLRW) metric, which takes the following form in Cartesian coordinates:

\begin{equation}
ds^2 = -dt^2 + a^2(t)\left(dx^2 + dy^2 + dz^2\right)
\end{equation}
here $a(t)$ denotes the scale factor of the Universe and $t$ represents the cosmic time.

\par\bigskip
A characteristic possibility in modified gravity is that the additional contribution $T_{\mu\nu}^{\mathrm{DE}}$ generates a sufficiently negative effective pressure, thereby offering a geometric explanation for the late-time accelerated expansion of the Universe. We formulate the $f(Q, C)$ cosmological equations in the coincident gauge, for which the affine connection vanishes identically, $\Gamma^{\alpha}_{\mu\nu}=0$. With this gauge choice, the relevant geometric quantities take the following form 

\begin{equation}
Q=-6H^2,\qquad
C=6(3H^2+\dot H),
\label{eq:flrwquantities}
\end{equation}

Substituting the preceding geometric expressions into the general field equations, Eq.~\ref{eq:fieldeq}, allows the cosmological evolution equations to be written in the standard Friedmann form, supplemented by an effective dark-energy contribution originating from the modified gravitational sector. Here, $\rho_m$ and $p_m$ denote the energy density and pressure of the matter component, respectively. The corresponding effective dark-energy density and pressure are defined by

\begin{strip}
\noindent
\rule{\dimexpr(\textwidth-\columnsep)/2\relax}{0.4pt}%
\rule{0.4pt}{5pt}\par
\vspace{6pt}

\begin{equation}
\begin{aligned}
\kappa \rho_{DE} &= -\frac{f}{2}+3H^2(1-2f_Q)+(9H^2+3\dot H)f_C-3H\dot f_C, \label{eq:rhoDE}
\end{aligned}
\end{equation}

\begin{equation}
\begin{aligned}
\kappa p_{DE} &= \frac{f}{2}-2\dot H(1-f_Q)-3H^2(1-2f_Q)+2\dot H f_Q - (9H^2+3\dot H)f_C+\ddot f_C. \label{eq:pDE}
\end{aligned}
\end{equation}

\vspace{6pt}
\noindent\hfill
\rule[-5pt]{0.4pt}{5.4pt}%
\rule{\dimexpr(\textwidth-\columnsep)/2\relax}{0.4pt}\par
\end{strip}

Here, $\dot{f}_C$ and $\ddot{f}_C$ denote the first and second cosmic-time derivatives of $f_C$. The resulting cosmological equations therefore generalize the standard FLRW equations of General Relativity by incorporating the additional geometric contributions arising from the dependence of the gravitational action on $Q$ and $C$. The first equation imposes the cosmological energy constraint, whereas the second governs the evolution of the cosmic expansion.\par\bigskip

A nonlinear gravitational Lagrangian extends the cosmological dynamics beyond those admitted by a linear form of $f(Q, C)$. Specifically, a power-law dependence on the non-metricity scalar renders the effective gravitational coupling sensitive to the underlying geometry, introducing additional contributions to the cosmic evolution. Retaining a linear term in $C$ preserves the boundary contribution linking the curvature and non-metricity formulations of gravity, although this term does not independently modify the bulk field equations when its coefficient is constant. Guided by previously explored nonlinear extensions of $f(Q)$ and $f(Q,C)$ gravity, we adopt the following phenomenological form

\begin{equation}
f(Q,C)=\gamma_1 Q^n+\gamma_2 C
\end{equation}

where $\gamma_1$, $\gamma_2$, and $n$ are the model parameters. The exponent $n$ characterizes the departure from a linear dependence on the non-metricity scalar $Q$, whereas $\gamma_1$ and $\gamma_2$ specify the relative weights of the non-metricity and boundary contributions in the gravitational action. This form provides a sufficiently general yet tractable framework for examining the modified cosmological dynamics and for confronting the model with observational data.\par\bigskip

\section{Parameterization of the deceleration parameter }

An appropriate parametrization of the deceleration parameter $q$ provides a useful basis for investigating the expansion history of the Universe. Previous studies have pursued both prescribed functional forms of $q$ and non-parametric reconstructions of its evolution. These approaches have been widely employed to examine cosmological questions concerning the initial singularity, the possibility of perpetual deceleration, the horizon problem, the Hubble tension, and related issues \cite{Banerjee2005, Cunha2008, Escamilla-Rivera2022}. Motivated by these investigations, we adopt a deceleration parameter whose cosmic-time evolution is specified through a linear dependence on the Hubble parameter $H$ \cite{Tiwari2017}.

\begin{equation}
q(z) = \alpha+\beta H,
\label{eq:qH}
\end{equation}

Here, $\alpha$ and $\beta$ are arbitrary constants. Specializing to $\alpha=-1$ in Eq.~\ref{eq:qH} and integrating the resulting evolution equation yields the following expression for the scale factor

\begin{equation}
a(t)=\exp\left(\frac{1}{\beta}\sqrt{2\beta t+b}\right),
\end{equation}

\par\bigskip
where $b$ is an integration constant. For the remainder of the analysis, we introduce the notation $F=\frac{1}{\beta}\sqrt{2\beta t+b}$. The adopted expansion law then yields the following expressions for the Hubble and deceleration parameters as functions of cosmic time $t$

\begin{equation}
H = \frac{1}{\beta F},
\qquad
q = -1 + \frac{1}{F}.
\end{equation}

\par\bigskip
The choice of the model parameter $\alpha=-1$ is motivated by the requirement of obtaining a time-dependent Hubble parameter that can describe a realistic transition from decelerated to accelerated cosmic expansion \cite{Garg2019, Sharma2019, Tiwari2018, Garg2021}. We therefore set $\alpha=-1$ to ensure that the deceleration parameter remains time-dependent. In contrast, for $\alpha\neq-1$ or $\alpha=0$, the resulting deceleration parameter becomes constant, with $q=-1$.\par\bigskip

The temporal evolution of the Hubble parameter is related to the deceleration parameter through $\dot{H}=-(1+q)H^{2}$. Employing this identity, we obtain the following expression for the Hubble parameter as a function of redshift

\begin{equation}
H(z)=\frac{1}{\beta(F_0-log(1+z))}
\end{equation}

Here, $F_0$ specifies the value of $F$ at the present epoch, corresponding to $z=0$. Furthermore, since the scale factor and redshift are related through

\begin{equation}
a(t)=\frac{1}{1+z},
\end{equation}

The corresponding relation between cosmic time and redshift takes the form

\begin{equation}
\frac{dt}{dz}=-\frac{1}{(1+z)H(z)}.
\end{equation}

Once the redshift dependence of the Hubble parameter is specified, this relation determines the corresponding evolution in cosmic time.

The same time–redshift relation allows the deceleration parameter to be expressed as a function of $z$. Performing this transformation yields

\begin{equation}
q(z) = -1+\frac{1}{F_0-\log(1+z)}.
\end{equation}
\par\bigskip

 Expressing these quantities in terms of the observable redshift provides a direct description of the cosmic expansion history. In particular, the redshift evolution of $q(z)$ makes it possible to trace the transition from the decelerating expansion of the early Universe to the accelerated phase at late times. The resulting field equations can then be specialized to the cosmological background to investigate the expansion history and the corresponding effective dark-energy behaviour.
 
\begin{strip}
\noindent
\rule{\dimexpr(\textwidth-\columnsep)/2\relax}{0.4pt}%
\rule{0.4pt}{5pt}\par
\vspace{6pt}

\begin{equation}
\begin{aligned}
\kappa\rho_{DE} &= 3H^2+\gamma_1(n-\tfrac12)(-6H^2)^n \label{eq:rhoDEfinal}
\end{aligned}
\end{equation}

\begin{equation}
\begin{aligned}
\kappa p_{DE} &= -\left[2\dot H+3H^2+\gamma_1(n-\tfrac12)(-6H^2)^n+ \frac{\gamma_1 n(2n-1)\dot H}{3}(-6)^n(H^2)^{n-1}\right]\label{eq:pDEfinal}
\end{aligned}
\end{equation}
\vspace{6pt}
\noindent\hfill
\rule[-5pt]{0.4pt}{5.4pt}%
\rule{\dimexpr(\textwidth-\columnsep)/2\relax}{0.4pt}\par
\end{strip}

Equations~\eqref{eq:rhoDEfinal}-\eqref{eq:pDEfinal} are the master expressions used throughout the remainder of the paper; they depend on the expansion history solely through $H(z)$ and $\dot H(z)$, which we now derive from the parametric ansatz.

\section{Observational Constraints }
\label{sec:data}

 To examine the observational consistency of the model, we derive the Hubble function $H(z)$ and constrain its parameters using cosmic chronometer (CC), baryon acoustic oscillation (BAO), Union3, and Pantheon+ Type Ia supernova observations. The likelihood functions are constructed from the corresponding $\chi^2$ statistics, and posterior parameter distributions are obtained through Markov Chain Monte Carlo (MCMC) sampling. The resulting constraints are used to determine the best-fit expansion history and evaluate the compatibility of the nonlinear $f(Q, C)$ model with observational data. The same analysis is performed for the spatially flat $\Lambda$CDM model, providing a consistent basis for comparison through parameter constraints.

\subsection{Cosmic Chronometer Dataset}

The cosmic chronometer (CC) method probes the expansion history of the Universe through the differential ageing of passively evolving galaxies. By comparing the ages of suitable galaxy populations observed at neighbouring redshifts, it estimates the elapsed cosmic time and hence the time–redshift relation. These populations therefore serve as cosmic clocks, enabling a direct determination of the Hubble expansion rate. Unlike probes based on luminosity or angular-diameter distances, this approach does not require a prescribed distance–redshift relation. The Hubble parameter is consequently expressed in terms of the variation of redshift with cosmic time as\par\bigskip

\begin{equation}
H(z)
=
-\frac{1}{1+z}\frac{dz}{dt},
\label{eq:cc}
\end{equation}
\par\bigskip

where $dz/dt$ denotes the variation of redshift with cosmic time, inferred from the differential ages of galaxies at neighbouring redshifts. The direct dependence on the time–redshift relation makes CC measurements a valuable probe of late-time cosmic expansion, providing constraints independent of the cosmic distance ladder.\par\bigskip

For the present analysis, we adopt a compilation of 31 cosmic chronometer measurements spanning $0.07\leq z\leq1.965$ \cite{Lamaaoune2025}. The observed Hubble parameter values, together with their corresponding measurement uncertainties, provide the observational constraints on the free parameters of the proposed cosmological model. To formulate the statistical comparison between theoretical predictions and observations, we treat the measurements as statistically independent and assume Gaussian-distributed errors. Under these assumptions, the chi-square statistic for the CC dataset takes the form

\begin{equation}
\chi_{\rm CC}^{2}
=
\sum_{i=1}^{31}
\left[
\frac{
H_{\rm obs}(z_i)
-
H_{\rm th}(z_i,\Theta)
}
{\sigma_i}
\right]^2
\end{equation}
\par\bigskip

where $H_{\rm obs}(z_i)$ is the measured Hubble parameter at redshift $z_i$, and $H_{\rm th}(z_i,\Theta)$ is its theoretical prediction for the parameter set $\Theta=(H_0,\beta)$. The quantity $\sigma_i$ denotes the uncertainty associated with the corresponding observation. The likelihood function for the CC dataset is therefore expressed as
\par\bigskip

\begin{equation}
\mathcal{L}_{\rm CC}
\propto
\exp
\left(
-\frac{\chi_{\rm CC}^{2}}{2}
\right),
\end{equation}
\par\bigskip

This likelihood underpins the parameter estimation performed in the present study. We sample the resulting posterior probability distribution with the affine-invariant ensemble Markov Chain Monte Carlo (MCMC) algorithm implemented in the \texttt{emcee} package. This procedure explores the joint parameter space, retaining correlations and degeneracies among the model parameters, and yields their marginalized posterior distributions and associated credible intervals. The constraints obtained from the CC dataset are subsequently employed to evaluate the observational consistency of the proposed model and to compare its predicted expansion history with cosmological measurements.\par\bigskip

\begin{figure}[H]
    \centering
      \includegraphics[width=0.99\linewidth]
      {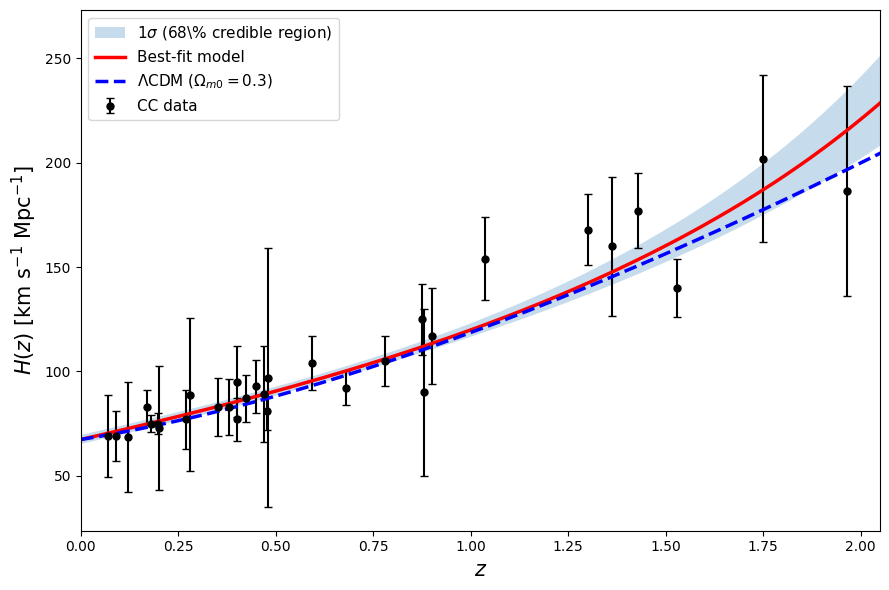}
    \caption{
Comparison of the best-fit linear model inferred from the joint analysis with the cosmic chronometer (CC) measurements. The red shaded band indicates the $1\sigma$ confidence interval, and the dashed blue curve shows the $\Lambda$CDM prediction.
}
    \label{fig: Joint}
\end{figure}

\begin{figure}[H]
    \centering
      \includegraphics[width=0.99\linewidth]
      {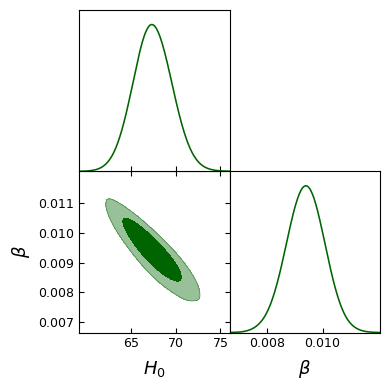}
    \caption{Joint constraints on the free model parameters inferred from the cosmic chronometer (CC) dataset, with contours delineating the corresponding $1\sigma$ and $2\sigma$ credible regions.}
    \label{fig: Hz_CC}
\end{figure}

Figure~\ref{fig: Joint} illustrates the Hubble parameter $H(z)$ reconstructed from cosmic chronometer (CC) observations, together with the corresponding $\Lambda$CDM prediction. Model parameters are constrained through Markov Chain Monte Carlo (MCMC) sampling, with their joint posterior distributions presented in Fig. \ref{fig: Hz_CC}. The inner and outer contours represent the $1\sigma$ and $2\sigma$ credible regions, respectively. The reconstructed expansion history shows broad agreement with $\Lambda$CDM at low redshifts, while deviations become more evident at higher redshifts. The marginalized parameter estimates and their associated uncertainties are summarized in the accompanying table.

\subsection{BAO Dataset}

Baryon acoustic oscillations (BAO) probe the geometry of cosmic expansion through a characteristic length scale preserved in the spatial distribution of matter. This feature originates from acoustic waves in the primordial baryon–photon plasma and appears as a preferred separation in galaxy and quasar clustering. Its well-defined physical scale enables measurements of the distance–redshift relation over a broad redshift range. The DESI Data Release 2 (DR2) analysis builds upon earlier BAO measurements, incorporating extensive galaxy and quasar observations across a wider redshift interval. \cite{Abdul2025}.\par\bigskip

For the present analysis, we adopt the official DESI DR2 BAO measurements used in the combined DESI BAO likelihood. This dataset comprises 13 correlated distance measurements spanning the effective redshift range $0.295\leq z_{\rm eff}\leq2.330$. Depending on the tracer population and redshift interval, the observables are reported as the transverse comoving distance $D_M/r_d$, the Hubble distance $D_H/r_d$, or the volume-averaged distance $D_V/r_d$. Their statistical correlations are accounted for using the full covariance matrix supplied with the DESI DR2 likelihood. Incorporating this covariance structure is essential for a statistically consistent inference of cosmological parameters from the combined BAO measurements.\par\bigskip

In the spatially flat FLRW geometry adopted here, the transverse comoving distance is determined by the Hubble expansion rate through the relation

\begin{equation}
D_M(z)=c\int_0^z\frac{dz'}{H(z')}
\end{equation}

where $c$ represents the speed of light. The corresponding Hubble distance is defined as 

\begin{equation}
D_H(z)=\frac{c}{H(z)}
\end{equation}

while the volume-averaged distance used to characterize the isotropic BAO measurement takes the form 

\begin{equation}
D_V(z)=\left[zD_M^2(z)D_H(z)\right]^{1/3}.
\end{equation}

The theoretical quantities compared with the DESI observations are therefore constructed as

\begin{equation}
\frac{D_M(z)}{r_d},\qquad
\frac{D_H(z)}{r_d},\qquad
\frac{D_V(z)}{r_d}
\end{equation} 

where $r_d$ denotes the comoving sound horizon evaluated at the baryon drag epoch. 

In our analysis, the sound-horizon scale is fixed to \(r_d=147.05\space{\rm Mpc}\), following the value adopted in our observational setup \cite{Planck:2018}. Since \(r_d\) is not treated as an additional free parameter, the BAO predictions are determined by the expansion history of the proposed cosmological model through \(H(z)\). Consequently, the DESI DR2 observations provide direct constraints on the model parameters that govern the redshift evolution of the Hubble parameter.\par\bigskip

The theoretical predictions are assessed against the DESI measurements using the full covariance matrix to account for correlations among the observables. The residual vector is defined as 

\begin{equation}
\Delta\mathbf{D}=
\mathbf{D}_{\rm th}-
\mathbf{D}_{\rm obs}
\end{equation}

The BAO chi-square is written as

\begin{equation}
\chi^2_{\rm BAO}
=\Delta\mathbf{D}^{\,T}
\mathbf{C}^{-1}
\Delta\mathbf{D}
\end{equation}

where $\mathbf{C}$ is the $13\times13$ covariance matrix encoding the uncertainties and correlations of the adopted DESI DR2 measurements. Under the assumption of Gaussian-distributed errors, the corresponding likelihood is expressed as 

\begin{equation}
\mathcal{L}_{\rm BAO}
\propto
\exp\left(-\frac{\chi^2_{\rm BAO}}{2}\right)
\end{equation}

The resulting likelihood is combined with the adopted priors to construct the posterior distribution, which is explored through Bayesian MCMC sampling to derive marginalized constraints on the free cosmological parameters. Comparing the theoretical predictions with the DESI DR2 measurements then provides an observational assessment of the expansion history implied by the proposed model.

\begin{figure}[H]
    \centering
      \includegraphics[width=0.99\linewidth]
      {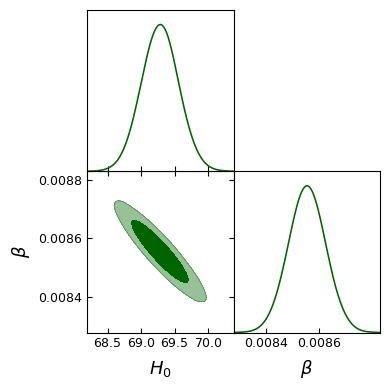}
    \caption{The likelihood contours presented in the figure illustrate the joint constraints on the free model parameters, with the corresponding $1\sigma$ and $2\sigma$ confidence regions obtained from the BAO dataset.}
    \label{fig: BAO}
\end{figure}

Figure~\ref{fig: BAO} presents the best-fit predictions of the proposed model alongside the DESI DR2 BAO measurements over the redshift range considered. The theoretical curves closely track the observations, supporting a consistent description of the measured BAO distance quantities. The shaded band represents the $1\sigma$ uncertainty obtained by propagating the parameter uncertainties inferred from the MCMC posterior distributions into the model predictions. This agreement indicates that the reconstructed expansion history adequately captures the late-time geometric information contained in the DESI DR2 BAO observations.

\subsection{Pantheon+SH0ES Dataset}

Type Ia supernovae (SNe Ia) serve as valuable distance indicators for tracing the late-time expansion history of the Universe. Empirical corrections to their observed luminosities enable their standardization, providing distance-modulus measurements across a broad redshift range. In this study, we employ the Pantheon+SH0ES compilation \cite{Brout:2022, Riess2022}, comprising 1701 light-curve measurements of spectroscopically confirmed SNe Ia spanning $0.001\leq z\leq2.26$. The compilation includes 77 measurements associated with supernovae in host galaxies whose distances are calibrated using Cepheid variables. These calibrators establish an absolute distance scale, enabling the Hubble constant to be constrained jointly with the remaining cosmological parameters.\par\bigskip

In a spatially flat FLRW spacetime, the comoving distance to a source at redshift $z$ is given by

\begin{equation}
D_C(z) = 
c\int_0^z\frac{dz'}{H(z')}
\end{equation}

where $c$ denotes the speed of light and $H(z)$ specifies the model-predicted expansion rate at redshift $z$. The corresponding luminosity distance is then obtained as

\begin{equation}
D_L(z)=(1+z)D_C(z)
\end{equation}

The theoretical prediction for the distance modulus takes the form

\begin{equation}
\mu_{\rm th}(z)=
5\log_{10}
\left[
\frac{D_L(z)}{\mathrm{Mpc}}
\right]+25.
\end{equation}
\par\bigskip
For the proposed linear model, the Hubble function follows from the adopted parametrization of the deceleration parameter. The corresponding luminosity distance is computed numerically from the integral involving $H(z)$ and evaluated at the redshifts of the observed supernovae. Specifically, we apply cumulative trapezoidal integration on a sufficiently fine redshift grid and interpolate the resulting distance–redshift relation to the individual supernova redshifts. This procedure yields the theoretical distance moduli used in constructing the likelihood.\par\bigskip

The Pantheon+SH0ES likelihood combines the corrected apparent magnitudes with the full covariance matrix supplied with the compilation, incorporating both statistical and systematic uncertainties in the supernova measurements. The SNe Ia absolute magnitude, $M$, is treated as a nuisance parameter and sampled jointly with the parameters governing the expansion history. Predictions for the Hubble-flow supernovae are calculated from the model luminosity distance, whereas those for the Cepheid-calibrated sample use the independently determined distances to their host galaxies. This formulation enables the supernova observations to constrain both the redshift dependence of the late-time expansion and the associated absolute distance scale.\par\bigskip

Agreement between the predicted and observed supernova magnitudes is assessed using the chi-square statistic associated with a Gaussian likelihood.

\begin{equation}
\chi^2=
\Delta\mathbf{m}^{,T}
\mathbf{C}^{-1}
\Delta\mathbf{m}
\end{equation}

where the residual vector is defined as

\begin{equation}
\Delta\mathbf{m}=
\mathbf{m}_{\rm obs}-
\mathbf{m}_{\rm model}
\end{equation}

and $\mathbf{C}$ represents the full $1701\times1701$ covariance matrix. To improve computational efficiency during MCMC sampling, its Cholesky factorization is computed once and reused in subsequent likelihood evaluations.\par\bigskip

The supernova likelihood is consequently written as

\begin{equation}
\mathcal{L}_{\rm SN}
\propto
\exp\left(-\frac{\chi_{\rm SN}^{2}}{2}\right).
\end{equation}
\par\bigskip
The posterior distributions of the model parameters are obtained by combining this likelihood with the adopted prior ranges within the MCMC framework. In particular, the parameters governing the linear deceleration-parameter parametrization, together with $H_0$ and the nuisance parameter $M$, are constrained simultaneously from the Pantheon+SH0ES observations.\par\bigskip

\begin{figure}[H]
    \centering
      \includegraphics[width=0.99\linewidth]
      {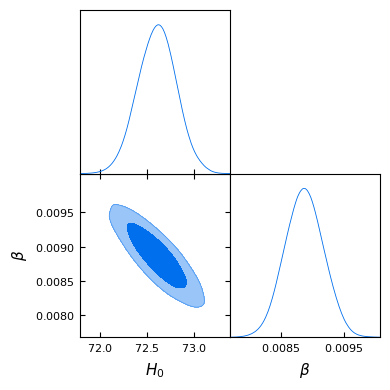}
    \caption{Joint constraints on the free model parameters derived from the Pantheon+SH0ES dataset, with contours enclosing the $1\sigma$ and $2\sigma$ posterior credible regions.}
   \label{fig: Pantheon}
\end{figure}

Figure~\ref{fig: Pantheon} presents the posterior constraints on the model parameters $(H_0,\beta)$ inferred from the Pantheon+SH0ES observations. The diagonal panels show the one-dimensional marginalized posterior distributions, whereas the off-diagonal panels depict the joint two-dimensional distributions and their associated parameter correlations. The contours delineate the $1\sigma$ and $2\sigma$ credible regions, illustrating both the uncertainties in the fitted parameters and the correlations between them.

\begin{figure}[H]
    \centering
      \includegraphics[width=0.99\linewidth]
      {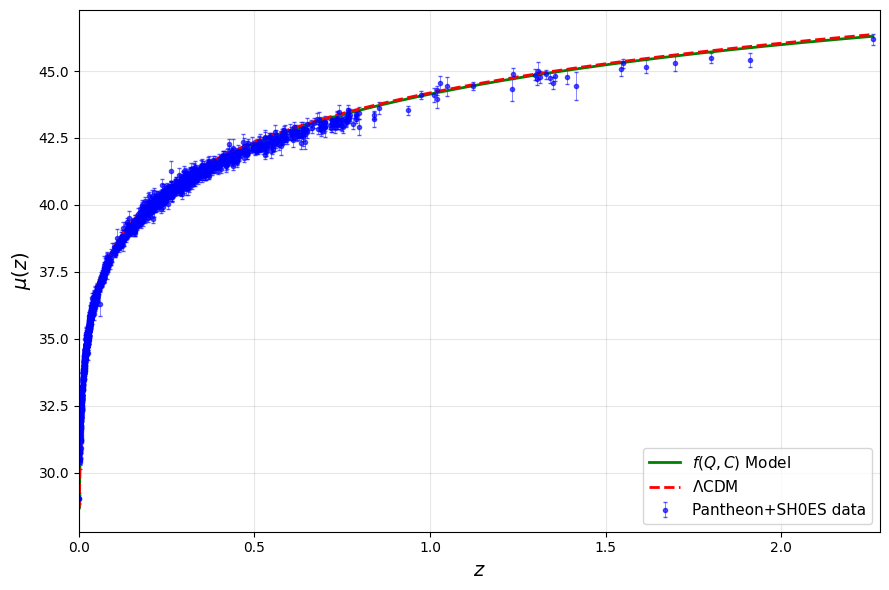}
    \caption{Error-bar representation of the 1701 Pantheon+SH0ES supernovae, shown together with the corresponding best-fit theoretical prediction.}
   \label{fig: Pantheon (error bar)}
\end{figure}

\subsection{Union3 Dataset}

To further assess the cosmological framework, we employ the Union3 Type Ia supernova compilation derived using the UNITY1.5 analysis \cite{Rubin:2025}. The full compilation comprises 2087 standardized Type Ia supernovae from 24 independent surveys, spanning a broad redshift interval that enables constraints on the late-time expansion history. Statistical and systematic effects associated with the supernova observations are incorporated through the hierarchical Bayesian framework of UNITY1.5.\par\bigskip

Rather than using the individual supernova measurements, we adopt the compressed Union3 likelihood provided by the collaboration. This release contains 22 binned distance-modulus measurements and the corresponding covariance information, covering approximately $0.05 \lesssim z \lesssim 2.26$. The use of the compressed likelihood substantially reduces the computational cost of the parameter estimation while retaining the information required for constraining the cosmological model.\par\bigskip

For a spatially flat FLRW background, the model-predicted luminosity distance is determined from the Hubble expansion rate through the relation

\begin{equation}
d_L(z)
=(1+z)c
\int_0^z
\frac{dz'}{H(z')}
\end{equation}
\par\bigskip
where $c$ denotes the speed of light and $H(z)$ is the Hubble parameter derived within the proposed cosmological model. The associated theoretical distance modulus follows from

\begin{equation}
\mu_{\rm th}(z)
= 5\log_{10}
\left(
\frac{d_L(z)}{\mathrm{Mpc}}
\right)
+25
\end{equation}
\par\bigskip
The theoretical distance moduli are fitted to the compressed Union3 data using a covariance-weighted chi-square statistic to determine the allowed ranges of the model parameters.

\begin{equation}
\chi^2_{\rm Union3}=
\Delta\boldsymbol{\mu}^{,T}
\mathbf{C}^{-1}
\Delta\boldsymbol{\mu}
\end{equation}

where the residual vector takes the form

\begin{equation}
\Delta\boldsymbol{\mu} =
\boldsymbol{\mu}_{\rm th}-
\boldsymbol{\mu}_{\rm obs}
\end{equation}

The matrix $\mathbf{C}^{-1}$ is the inverse covariance matrix for the 22 compressed distance-modulus measurements and specifies their statistical weighting in the analysis.

\begin{figure}[H]
    \centering
      \includegraphics[width=0.99\linewidth]
      {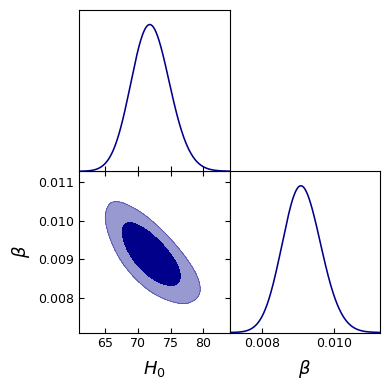}
    \caption{Joint posterior constraints on the free model parameters inferred from the Union3 Type Ia supernova dataset. The contours delineate the corresponding $1\sigma$ and $2\sigma$ credible regions.}
    \label{fig: Union3}
\end{figure} 

\subsection{Combined Analysis}

Joint constraints on the proposed model are derived by incorporating the cosmic chronometer (CC), DESI DR2 BAO, Pantheon+SH0ES, and Union3 observations into a common likelihood analysis. Under the assumption of statistical independence among these datasets, the combined likelihood factorizes into the product of their individual likelihoods. The total chi-square is therefore expressed as the sum of the contributions from the respective observational probes \par\bigskip

\begin{equation}
\chi_{\rm Combined}^{2}
=
\chi_{\rm CC}^{2}
+
\chi_{\rm BAO}^{2}
+
\chi_{\rm Pantheon+SH0ES}^{2}
+
\chi_{\rm Union3}^{2}
\end{equation}

\begin{figure}[H]
    \centering
      \includegraphics[width=0.99\linewidth]
      {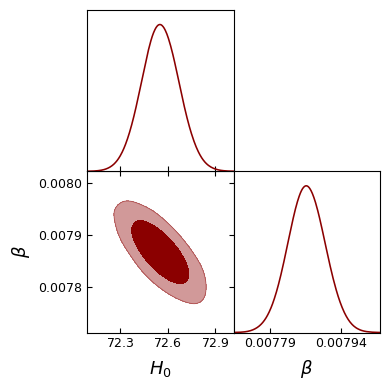}
   \caption{Joint posterior constraints on the free model parameters from the combined CC, DESI DR2 BAO, Pantheon+SH0ES, and Union3 analysis. The contours enclose the $1\sigma$ and $2\sigma$ credible regions.}
\label{fig: Combined}
\end{figure}

The combined analysis brings together complementary constraints from direct measurements of the expansion rate, BAO distance scales, and Type Ia supernova luminosity distances. By incorporating all four datasets simultaneously, the joint posterior characterizes the allowed parameter space of the model. Figure~\ref{fig: Combined} presents the marginalized posterior distributions and the associated two-dimensional credible contours, while Table~\ref{tab:mcmc_results} summarizes the inferred best-fit parameter values and their uncertainties.\par\bigskip

\begin{table*}[ht]
\centering
\caption{Best-fit model parameters inferred from the CC, DESI DR2 BAO, Pantheon+SH0ES, and Union3 datasets, together with their associated $1\sigma$ uncertainties\\.}

\label{tab:mcmc_results}
\begin{tabular}{lccc}
\toprule
Dataset & $H_0$ (km s$^{-1}$ Mpc$^{-1}$)  & $\beta$ \\
\midrule
CC &
$67.3744^{+3.6733}_{-3.9121}$ &
$0.6318^{+0.2025}_{-0.1438}$ \\
\\
Union3 &
$71.9340^{+2.9695 }_{-2.8568}$ &
$0.0091^{+0.0005}_{-0.0005}$ \\
\\
BAO &
$69.2821^{+0.2762}_{-0.2781}$ &
$0.0086^{+0.0001}_{-0.0001}$ \\
\\
Pantheon+SH0ES &
$72.5959^{+0.2012}_{-0.1994}$ &
$0.0089 ^{+0.0003}_{-0.0003}$ \\
\\
Combined &
$\mathbf{72.55242}^{+0.11903}_{-0.11752}$ &
$\mathbf{0.0079}^{+0.0003}_{-0.0003}$ \\
\bottomrule
\end{tabular}

\end{table*}

\section{Cosmographic Diagnostics}
\label{sec:cosmography}

\subsection{Deceleration parameter}

The deceleration parameter provides a fundamental kinematic measure of the acceleration or deceleration of cosmic expansion and is defined by

\begin{equation}
q=-\frac{\ddot{a}}{aH^{2}}
\end{equation}

where $a(t)$ is the cosmic scale factor and $H=\dot{a}/a$ denotes the Hubble parameter, with overdots representing derivatives with respect to cosmic time. Expressed in terms of redshift $z$, this definition takes the form

\begin{equation}
q(z)=-1+\left(\frac{1+z}{H(z)}\right)\frac{dH(z)}{dz}.
\end{equation}
\par\bigskip
In the present model, the deceleration parameter is calculated from the parametrically reconstructed Hubble function $H(z)$, enabling the different expansion regimes and the transitions between them to be characterized directly. Positive values, $q(z)>0$, signify decelerated expansion, as expected when the attractive gravitational influence of the cosmic matter content dominates. Conversely, $q(z)<0$ indicates accelerated expansion, associated here with the growing influence of the effective dark-energy sector. The value $q(z)=0$ separates these two regimes and identifies the transition when the parameter changes sign. A sustained value of $q=-1$ corresponds to the de Sitter limit of exponential expansion, whereas $q<-1$ characterizes a super-accelerating, or phantom-like, regime. Tracking the redshift evolution of $q(z)$ therefore provides a direct means of examining the expansion history and evaluating whether the proposed model reproduces the observed transition from an earlier decelerating phase to the present epoch of acceleration.\par\bigskip

\begin{figure}[H]
    \centering
      \includegraphics[width=0.99\linewidth]
      {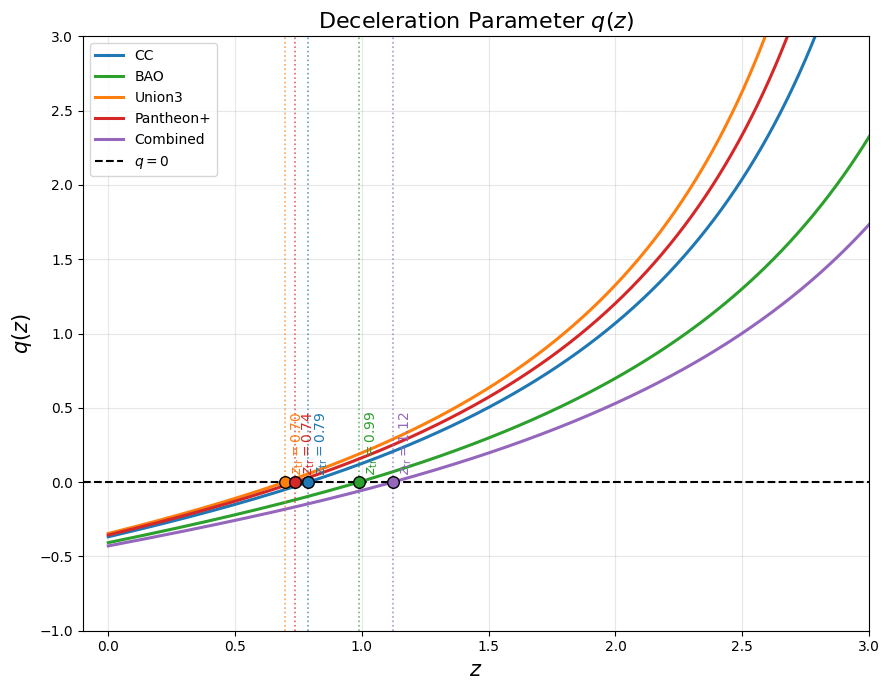}
    \caption{Evolution of the deceleration parameter $q(z)$ with redshift for the proposed linear parametrization, evaluated using the best-fit parameters inferred from the individual CC, DESI DR2 BAO, Pantheon+SH0ES, and Union3 datasets and their combined analysis.}
    \label{fig:Deceleration parameter}
\end{figure}
\par\bigskip
Figure~\ref{fig:Deceleration parameter} traces the evolution of the deceleration parameter from an earlier decelerating phase, where $q(z)>0$, to late-time accelerated expansion, where $q(z)<0$. The observationally constrained parameters yield transition redshifts of $z_{\rm tr}\simeq0.70$, $0.74$, $0.79$, $0.99$, and $1.12$ for Union3, Pantheon+, CC, DESI BAO, and the combined analysis, respectively. The combined constraints therefore favour an earlier onset of cosmic acceleration than the individual datasets. In the same order, the present-day values are $q_0\simeq-0.34$, $-0.35$, $-0.36$, $-0.40$, and $-0.42$, consistently indicating accelerated expansion at the current epoch. The subsequent evolution of $q(z)$ toward $-1$ suggests an asymptotic approach to de~Sitter-like kinematic behaviour, consistent with an increasing influence of the effective dark-energy sector on the expansion dynamics \cite{Cunha_2009, Nair_2012}.

\subsection{Equation of state}

The equation-of-state (EoS) parameter characterizes the relationship between pressure and energy density, providing insight into the dynamical behaviour of the cosmic components and their roles in successive stages of cosmic evolution. For the spatially flat FLRW background considered here, the effective dark-energy EoS parameter is defined as\par\bigskip

\begin{equation}
\omega_{\mathrm{DE}}(z)=\frac{p_{\mathrm{DE}}(z)}{\rho_{\mathrm{DE}}(z)}.
\end{equation}

Within the adopted linear parametrization of the deceleration parameter, we reconstruct this evolution using the best-fit parameters inferred from the CC, DESI DR2 BAO, Pantheon+SH0ES, and Union3 datasets, together with those obtained from their combined analysis.\par\bigskip

The different ranges of $\omega$ characterize distinct dynamical regimes. The values $\omega=1/3$ and $\omega=0$ correspond to radiation and matter-dominated expansion, respectively, while $\omega=-1$ represents the de Sitter limit associated with a cosmological constant. The accelerated phase requires $\omega<-1/3$. In particular, the range $-1<\omega <-1/3$ corresponds to a quintessence-like regime \cite{Tsujikawa_2013}, whereas $\omega<-1$ represents a phantom-like behaviour. These regimes provide a convenient way of interpreting the late-time dynamics obtained from the linear parametrization.\par\bigskip

The reconstructed evolution of $\omega(z)$ provides insight into the dark-energy dynamics accompanying the transition from earlier deceleration to the present accelerated expansion. Results inferred from the different observational combinations are compared with the $\Lambda$CDM benchmark, in which the dark-energy equation of state remains fixed at $\omega_\Lambda=-1$. This comparison assesses how closely the late-time behaviour of the linear model approaches the cosmological-constant regime and establishes the extent to which the observational datasets constrain its dynamical evolution.\par\bigskip

\begin{figure}[H]
    \centering
      \includegraphics[width=0.99\linewidth]
      {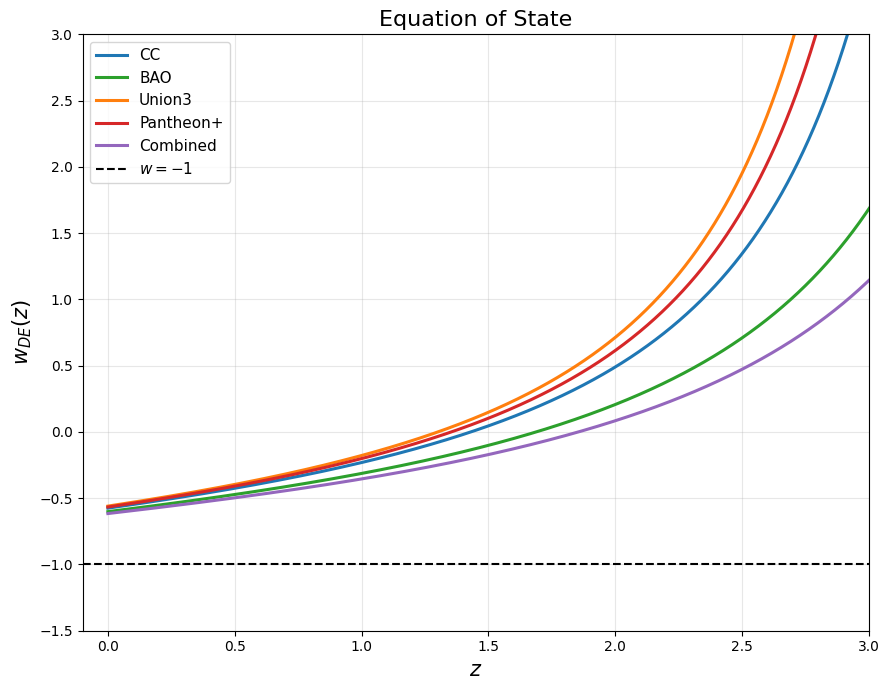}
    \caption{Redshift evolution of the effective dark-energy equation-of-state parameter $w_{\mathrm{DE}}(z)$ for the proposed linear parametrization, reconstructed from the best-fit parameters inferred from the individual CC, DESI DR2 BAO, Pantheon+SH0ES, and Union3 datasets and their combined analysis.}
    \label{fig:Eos}
\end{figure}

Figure~\ref{fig:Eos} presents the redshift dependence of the equation-of-state parameter reconstructed using the best-fit parameters of the linear model for the CC, Union3, DESI DR2 BAO, Pantheon+SH0ES, and combined analyses, with the corresponding $\Lambda$CDM prediction included for comparison. At the present epoch, $z=0$, the inferred values are $\omega_{\mathrm{DE}}(0)=-0.5751$, $-0.6015$, $-0.5598$, $-0.5665$, and $-0.6159$, respectively. Each estimate falls within $-1<\omega_{\mathrm{DE}}<-1/3$, consistently placing the reconstructed effective dark-energy component in the quintessence-like regime at the current epoch.\par\bigskip

\subsection{Statefinder diagnostics}

The late-time acceleration of cosmic expansion is commonly attributed to dark energy (DE), although its physical origin and defining properties remain unresolved. The diversity of proposed DE models motivates the use of robust diagnostics capable of distinguishing their qualitative behaviour and quantitative predictions. To this end, Sahni et al. \cite{Sahni_2003} introduced the statefinder diagnostic, which characterizes cosmic evolution geometrically through the parameter pair $(r,s)$. These parameters are defined as

\begin{equation}
r(z) = 2q^2+q+(1+z)\frac{dq}{dz} \label{eq:rstate}
\end{equation}

\begin{equation}
s(z) = \frac{r-1}{3\left(q-\tfrac12\right)} \label{eq:sstate}
\end{equation}

\begin{figure}[H]
    \centering
      \includegraphics[width=0.99\linewidth]
      {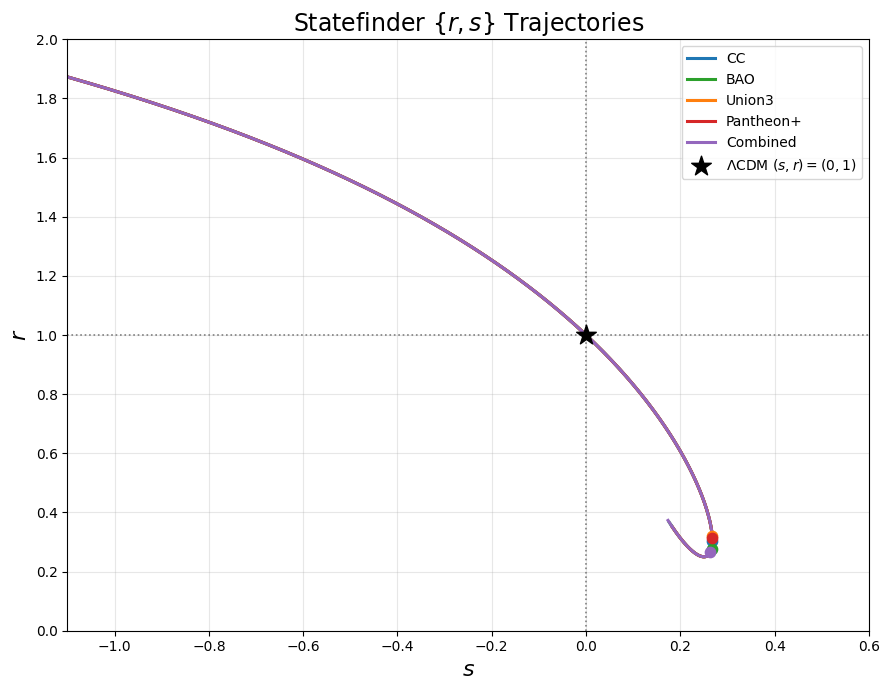}
    \caption{Statefinder trajectories in the $(r,s)$ plane for the proposed linear parametrization, reconstructed from the best-fit parameters inferred from the individual CC, DESI DR2 BAO, Pantheon+SH0ES, and Union3 datasets and their combined analysis.}
    \label{fig:Statefinder}
\end{figure}

The statefinder pair offers a geometric means of distinguishing dark-energy scenarios through their trajectories in the $(r,s)$ plane. The $\Lambda$CDM model corresponds to the fixed point $(r,s)=(1,0)$, while the regions $r>1$, $s<0$ and $r<1$, $s>0$ are commonly associated with Chaplygin-gas-like and quintessence-like behaviour, respectively. Figure~\ref{fig:Statefinder} presents the statefinder trajectories reconstructed from the CC, DESI DR2 BAO, Pantheon+SH0ES, Union3, and combined datasets. The trajectories remain within the region $r>1$ and $s<0$, exhibiting a geometric signature characteristic of Chaplygin-gas-like evolution. Their gradual approach to $(1,0)$ further indicates convergence toward the $\Lambda$CDM statefinder values at late times. This diagnostic therefore characterizes the model’s dynamical evolution and its proximity to the standard cosmological scenario at the level of the statefinder parameters.\par\bigskip

\subsection{ Om diagnostic}

The $Om(z)$ diagnostic offers a simple geometric means of assessing departures from $\Lambda$CDM and distinguishing the expansion histories associated with different dark-energy (DE) models \cite{Sahni_2008}. Unlike higher-order diagnostics such as the statefinder pair, it requires only the Hubble parameter $H(z)$ and the redshift $z$, making its evaluation comparatively straightforward. For a spatially flat Universe, it is defined by\par\bigskip

\begin{equation}
Om(z)=\frac{\left[H(z)/H_{0}\right]^{2}-1}{(1+z)^{3}-1}.
\end{equation}

\begin{figure}[H]
    \centering
      \includegraphics[width=0.99\linewidth]
      {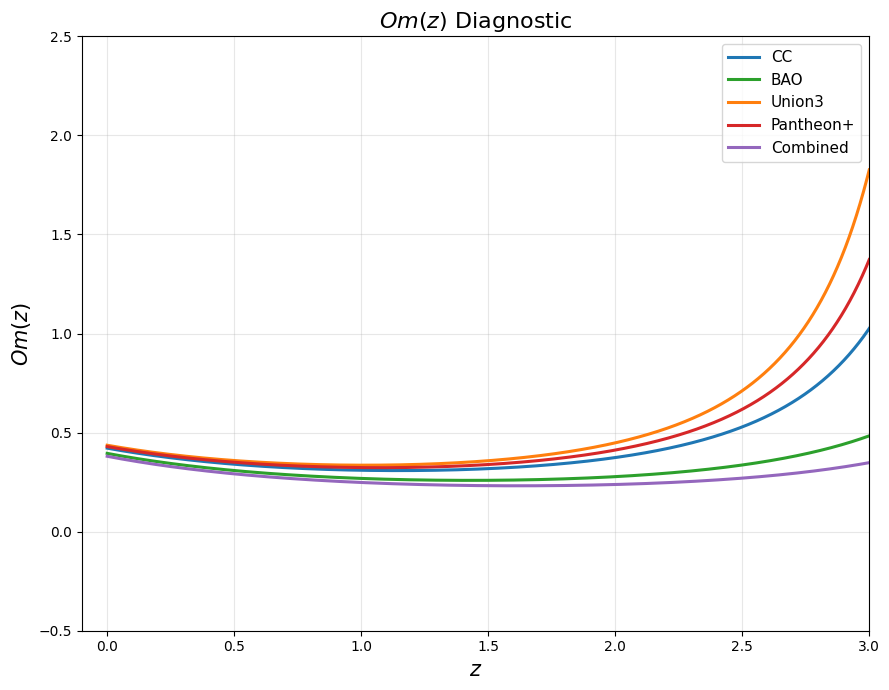}
    \caption{Evolution of the $Om(z)$ diagnostic with redshift for the proposed linear parametrization, evaluated using the best-fit parameters inferred from the individual CC, DESI DR2 BAO, Pantheon+SH0ES, and Union3 datasets and their combined analysis.}
    \label{fig:Om}
\end{figure}

Within the standard interpretation of this diagnostic, a negative slope is associated with quintessence-like behaviour, $\omega>-1$, whereas a positive slope is associated with phantom-like behaviour, $\omega<-1$. Figure~\ref{fig:Om} displays the reconstructed $Om(z)$ trajectories for the CC, DESI DR2 BAO, Pantheon+SH0ES, Union3, and combined analyses. All trajectories exhibit a negative slope over the plotted redshift interval, suggesting a quintessence-like effective behaviour within this diagnostic interpretation.
\par\bigskip

\section{Discussion and Conclusions}

In this study, we have examined late-time cosmic expansion in \(f(Q, C)\) gravity by adopting a linear dependence of the deceleration parameter on the Hubble rate, \(q=-1+\beta H\). This assumption leads to the analytical Hubble function \(H(z)=\frac{1}{\beta[F_0-\ln(1+z)]}\), which provides a consistent basis for evaluating the expansion history and the associated cosmological diagnostics. Here, \(F=\frac{1}{\beta}\sqrt{2\beta t+b}\), with \(F_0\) denoting its present-day value. The gravitational action combines a nonlinear function of the non-metricity scalar \(Q\) with a linear boundary term whose coefficient is constant. Since this boundary contribution leaves the bulk field equations unchanged, the modifications to the gravitational dynamics originate entirely from the nonlinear dependence on \(Q\).\par\bigskip

The free parameters of the constructed cosmological model are constrained using a combination of four complementary observational data sets: Cosmic Chronometer (CC),  BAO measurements, Pantheon+SH0ES, and Union3. A Bayesian statistical framework is considered, with the parameter space explored through an affine-invariant Markov Chain Monte Carlo (MCMC) approach. The constraints derived from individual datasets exhibit good mutual agreement, indicating the stability of the model under different observational probes. Furthermore, the joint analysis significantly improves the precision of the parameter estimates by exploiting the complementary information contained in the datasets. The resulting best-fit parameter values reproduce the observed late-time cosmic expansion history with high accuracy across the stipulated redshift range, thereby supporting the viability and observational reliability of the proposed model.\par\bigskip

The reconstructed cosmic expansion within this framework finds a robust geometric interpretation through the effective energy density ($ \ rho _ {DE} $), pressure $(p_{DE})$, and the time-dependent equation of state parameter$(\omega_{DE})$, all derived self-consistently via the modified gravitational field equations. While traditional kinematic variables such as the Hubble parameter$(H(z))$  and deceleration parameter$q(z)$ map the raw expansion history, they remain insufficient for breaking the degeneracy between competing dark energy models. The behavior of $\ rho _ {DE}$, $(p_{DE})$, and $(\omega_{DE})$ complements this kinematic dataset by proving a dynamic profile of the cosmic fluid. Crucially, the evolution of the effective EoS parameter offers insight into the possible emergence of dark-energy-like behavior without requiring an independent phenomenological prescription for the dark-energy equation of state. \par\bigskip
To further isolate the subtle features of this expansion history—which are frequently obscured when analyzing $H(z)$ alone—we employ higher-order geometric diagnostics: the statefinder parameter pair$(r,s)$ and the $Om(z)$ diagnostic. The statefinder parameters, which depend on the third derivative of the scale factor $\dddot{a}$, act as sensitive probes of deviations from standard cosmological models. An asymptotic or local approach to the statefinder fixed point $(r,s)=(1,0)$ indicates a statistical and geometric convergence with the baseline $\Lambda$ CDM model. However, we emphasize that matching these diagnostic values does not imply physical and structural equivalence between the underlying gravitational frameworks; distinct geometric actions can yield identical expansion trajectories. Because the expansion history in this work is prescribed phenomenologically and subsequently mapped onto the gravitational field equations, the utilization of background observational data (such as Type Ia Supernovae, Cosmic Microwave Background shift parameters, and Baryon Acoustic Oscillations) acts to constrain the parameters of this specific kinematic prescription. It does not independently validate or establish the uniqueness of the chosen gravitational action. Furthermore, because the boundary term $C$ enters the action linearly, its contribution integrates out as a total derivative in the field equations, rendering the bulk cosmic dynamics identical to those of pure symmetric teleparallel $f(Q)$ gravity. As a result, the present analysis explores a restricted, linear sector of $f(Q, C)$ gravity and does not probe or constrain the richer, non-linear dynamical degrees of freedom that would naturally arise from a non-linear dependence on the boundary term $C$.

\section*{Data Availability Statement}
No new data were generated in this study; the datasets used are publicly available from the sources.

\balance
\bibliographystyle{unsrtnat}
\bibliography{reference}

\end{document}